\documentclass[fleqn,usenatbib,letters]{mnras}

\usepackage{newtxtext,newtxmath}

\usepackage[T1]{fontenc}
\usepackage{ae,aecompl}
\usepackage{xcolor}

\usepackage{graphicx}	
\usepackage{amsmath}	
\usepackage{amssymb}	

\title[SNIa and radio mini-halos]{Role of intracluster supernovae in radio mini-halos in galaxy clusters}

\author[A. Omar]{
A. Omar\thanks{E-mail: aomar@aries.res.in (AO)}
\\
Aryabhatta Research Institute of observational-sciences, Manora Peak, Nainital, 263001, India\\
}

\begin{document}
\maketitle

\begin{abstract}
A possibility of generating a population of cosmic-ray particles accelerated in supernovae type-Ia (SNIa) remnants in the intracluster medium (ICM) is discussed. The presently constrained host-less SNIa rates in the clusters are found to be sufficient to fill a few hundred kpc region with cosmic-ray electrons within their typical synchrotron life-time of 100 Myr. The SNIa have already been considered potential sources of excess Fe abundance in cool-core clusters, distributed heating and turbulence in ICM. A good fraction of total radio power from mini-halos can be sourced from the SNIa energy deposited in the ICM with required energy conversion efficiency $\le1$ per cent. The radio power estimated from low Mach-number shock acceleration in SNIa remnants is consistent with the observations within the uncertainties in the estimates. Some observational properties of the radio mini-halos are broadly consistent with the SNIa scenario. It is also speculated that radio powers and possibly detections of mini-halos are linked to star formation and merger histories of the clusters. 
\end{abstract}

\begin{keywords}
galaxies: clusters: intracluster medium -- supernovae: general -- radio continuum: general -- cosmic rays
\end{keywords}



\section{Introduction}

Some nearby ($z<0.5$) dynamically-relaxed clusters of galaxies are known to have diffuse synchrotron radio emission over $50-300$ kpc radius in the intra-cluster medium (ICM) around the brightest cluster galaxy (BCG) (e.g., \citealt{Feretti2012}; \citealt{Kale13}; \citealt{Kale15}; \citealt{Gia14, Gia17}). This low surface brightness ($\le$1~$\mu$Jy arcsec$^{-2}$ at 1.4 GHz) radio emission over a few hundred kpc extent is termed as the radio mini-halos (RMH; \citealt{Feretti2012}). The RMH have radio powers ($\nu P_{\nu}$) in between 10$^{39}$ - 10$^{42}$ erg~s$^{-1}$ (Bravi, Gitti \& Brunetti 2016). The average magnetic field strength in the ICM within the extent of RMH, derived using the equipartition assumption, is typically found $\le1\mu$G \citep{Kale13}. The RMH have been detected only in the clusters, which show a cool core, however, not all the observed cool-core clusters are detected with RMH. Some massive merging clusters also show radio relics at the cluster periphery and giant radio halos over larger scales in the clusters (e.g., see \citealt{Cassano,george,Gia17,vanWeeren}). As all of these diffuse radio emissions are non-thermal in nature, a common physical process is often invoked to explain their origins. Although, the physical processes powering the diffuse radio emission in the clusters are not well identified, it is believed that re-accelerations of suprathermal/mildly-relativistic electrons present in the cluster volume in some physical processes like turbulence or shocks give rise to a population of ultra-relativistic cosmic-ray electrons (CRe) which in turn generate synchrotron radio emission in the presence of magnetic field in the clusters. The strong shocks in the ICM can be generated in cluster-cluster merger, structure formation, AGN and SNe explosions.

The detection of synchrotron radio emission requires that the particle acceleration is presently ongoing or happened very recently within about 100 Myr \citep{Sarazin}. The models based on the re-acceleration of CRe are termed as the leptonic models. An alternate scenario termed as the hadronic model predicts that the inelastic collisions between the relativistic protons (CRp) and the thermal protons can produce CRe \citep{Den,Pfro}. The radio relics seen primarily at the peripheries of some clusters are associated with particle accelerations taking place in strong structure formation and cluster-cluster merger shocks (e.g., \citealt{Sarazin}; \citealt{Roettiger}). The emission from the radio halos requires a more distributed source of cosmic-ray (CR) particle acceleration or production. In context of the leptonic models, effective generation of turbulence in the ICM is a key component. As there are no recent major mergers, it is hypothesised that turbulence in the cool-cores can be generated by some processes such as compression of the ICM in cooling flows \citep{Gitti02, Gitti04} and gas-sloshing in the potential well of the cluster initiated by a minor merger \citep{Mazzotta, ZuHone, Gia14}. A requirement of re-acceleration of CRe over the cluster-wide scales comes from the fact that the synchrotron cooling time ($\sim$100 Myr) of the ultra-relativistic CRe is shorter than the required diffusion time ($\ge1$ Gyr) for the full extents of the detected radio emission in the clusters. Therefore, the central sources of CRe such as active galactic nuclei (AGN) or core-collapse supernovae (SNcc) in nuclear star-bursts in the galaxies cannot effectively supply these electrons over several 100 kpc-scale unless some in-situ production or re-acceleration of these electrons takes place in the ICM. 

In the interstellar medium (ISM), much of the particle acceleration is believed to be taking place in the shocks generated from the blast waves in supernovae (SNe) remnants in diffusive shock acceleration (DSA) process (see \citealt{Reynolds} for a review). The energy density of the resulting CRp is expected to be much higher than that of CRe in the DSA models. The DSA models alone are generally not able to explain all the observed properties of the diffuse radio emission in the ICM and the 2-step process combining DSA and re-acceleration in turbulence is invoked (see e.g., \citealt{brunetti,brunetti14,Pinzke}). The re-acceleration models require a pre-existing population of supra-thermal/mildly-relativistic fossil electrons in the ICM \citep{Pinzke}. The present-epoch populations of the fossil electrons required in re-acceleration models and protons required in hadronic models are assumed to have originated in some shocks associated with cluster merger or accretion in the ICM or injected into the ICM from radio galaxies or SNe-driven galactic winds over a few Gyr in the past. 

The efficiency of acceleration of particles to the ultra-relativistic energies in the ICM is highly uncertain. The relatively low sonic Mach numbers ($\mathcal{M}$) in hot medium make electron acceleration efficiency in the DSA process low in the ICM \citep{Dorfi96,Hoeft,Kang13}. The diffuse radio emission associated with the relics, detected at the peripheries in some merging galaxy clusters, is nevertheless explained via acceleration of CR particles in such low Mach-number shocks (e.g., \citealt{Kang, vanWeeren}). The low-$\mathcal{M}$ shocks have been routinely identified in the galaxy clusters via X-ray and radio emission {\color{blue}(Colafrancesco, Marchegiani \& Paulo 2017} and references therein). The observed spectral indices of the RMH are typically found between 1.2 and 1.6 \citep{Gia14, Kale15}. Considering that $\mathcal{M}$ is linked to the spectral index (see e.g., \citealt{col}), $\mathcal{M}$ corresponding to spectral index 1.2 - 1.6 will be between 2 and 3. Therefore, the radio emission from RMH is likely to be produced in low-$\mathcal{M}$ shocks in the ICM. The shocks described in the ICM such as the merger shocks have typical velocities of $\sim$1000 km~s$^{-1}$, similar to the free-expansion velocity of the SNe ejecta. 

In this paper, we discuss a possibility for the remnants of intra-cluster (IC) SNe of type Ia (SNIa) providing a naturally distributed source of CR particles produced in the SNe shocks. The motivation for the discussions presented in this paper comes from high rates of IC-SNIa predicted based on the observations of host-less SNe in some clusters.

\section{Importance of SNI\lowercase{a} in ISM and ICM}

The SNIa is a cataclysmic explosion of a low-mass star, very likely a white dwarf, in a binary system. The explosion is believed to take place when the white dwarf accretes mass up to the Chandrashekhar limit from the companion star and undergoes a runaway thermonuclear fusion. In contrast, the SNcc events, which mark the end of short-lived ($<10$ Myr) massive ($>8$ M$_{\odot}$) stars, take place mostly during a star-formation phase in galaxies. Being associated with the low-mass stars, SNIa can occur over a longer period of time after an initial burst of star formation. In terms of the energetics, the ejecta of either SNIa or SNcc carry kinetic energy of $\sim$10$^{51}$ erg. The SNIa and their remnants are believed to be playing important roles in the metal enrichment of ISM with heavy elements (e.g., \citealt{Wiersma}), galaxy evolution processes (e.g., \citealt{Powell}), and acceleration of cosmic rays \citep{Reynolds}. In ICM, the SNIa are considered an important source of heating and also responsible for iron (Fe) and other metals enrichment (e.g., Zaritsky, Gonzalez \& Zabludoff 2004; \citealt{Tang}). 

The galaxy clusters are broadly divided in two categories, namely, cool-core (CC) and non-cool core clusters. The inner regions of the CC clusters show drop in the temperature of the X-ray emitting gas by about one-third of that in the outer regions (e.g., \citealt{Hudson, sanderson}). The cooling time of the thermal gas in the cores of the clusters is expected to be very short ($<1$ Gyr) compared to the Hubble time \citep{Sarazin}. Therefore, in absence of a heating source in the ICM, the gas temperature profiles are expected to be very steep and the temperature in the central region is expected to drop by orders of magnitude. As such steep thermal profiles are never detected in the ICM, it is believed that the clusters also have significant heating sources (see e.g., \citealt{roychow} and references therein). Some of the heating sources are proposed as AGN outflows, thermal conduction from outer regions of the clusters, cluster-cluster merger, gas sloshing, and IC-SNe explosions. As the SNIa events are randomly distributed in the IC volume, SNIa are considered efficient sources of distributed heating and turbulence over large volumes in low density hot medium typical of that in the ICM (see e.g., \citealt{Valdarnini, Domainko04, Tang}). 

The SNe explosions are the major source of metal enrichment of the ICM. The SNIa become far more important in the ICM to explain the observed Fe and other metal (e.g., O, Ne, Mg, SI, S, Ar) abundances in the clusters. The CC clusters show enhancements of Fe and other metals abundances in the inner $\sim$100 kpc region (e.g., \citealt{Finoguenov}; \citealt{Degrandi}). While O, Ne, and Mg are produced almost entirely in the SNcc, Fe comes mostly from SNIa and other elements such as Si, S, and Ar can come from both SNIa and SNcc (e.g., \citealt{Nomoto}). The analyses of relative abundance profiles in the CC clusters require dominance of SNIa in the central regions and total enrichment over entire ICM being comparable from both SNIa and SNcc (e.g. \citealt{Dupke, Mernier}). \citet{Zaritsky} and \citet{Domainko04} estimated a few times $10^{9}$ IC-SNIa required to match the observed Fe abundance excess in the CC clusters. It is not entirely clear what makes only the CC clusters to have excess Fe abundance, although, links to the stellar content of BCG \citep{Mernier}, stripping of matter from the infalling galaxies \citep{Domainko06}  and  star formation history since the last major merger \citep{Degrandi} are suggested. 

Classically, the remnants of SNIa have been considered less important than the remnants of SNcc to explain the radio continuum emission, dominantly of the synchrotron origin, from the ISM in star-forming galaxies \citep{Condon}. The synchrotron radio emission can typically last for $\le$100 Myr after the cessation of massive star formation in galaxies. The SNe events in the star-forming galaxies are primarily of the SNcc types. An analysis predicts SNIa events at a level of 15 - 20 per cent of the total SNe in the first 100 Myr since the initiation of star formation in galaxies \citep{Bonaparte}. At later epochs, the SNIa events are expected to dominate over the SNcc events. \citet{Condon} considered that the SNIa remnants might contribute to the radio emission in galaxies with very low star formation rates. \citet{Volk}  discussed a possibility of acceleration of CR particles in SNIa in early-type galaxies in the clusters. As both types of SNe remnants can accelerate cosmic particles to the relativistic speeds (see e.g., \citealt{Reynolds}), the role of SNIa remnants in particle acceleration can not be neglected, particularly at late epochs after the cessation of star-burst or in galaxies with very low star formation rates. The detection of the synchrotron radio emission at late epochs or from galaxies with very low star formation rates will, however, depends also on the magnetic field strengths, which may not be sufficient in such galaxies to generate detectable radio emission at the current sensitivities of the radio telescopes. 

\section{SNI\lowercase{a} and radio mini-halos}

The coincidences of RMH being found exclusively in the CC clusters, excess central Fe abundance very likely due to SNIa in the CC clusters, and SNIa being a potential source of distributed turbulence and heating in the ICM, motivate to find a possible link between SNIa and RMH. The detections of IC-SNIa  strengthen this possibility. A good number of SNe almost all of them as SNIa type are found to be host-less in the ICM, not associated with a galaxy \citep{Gal-yam, Sand, Dilday, sand11}. It is found that about 30 per cent of total SNIa in groups are host-less \citep{Mcgee} and about 10 per cent of total SNIa in rich clusters are host-less  \citep{Dilday}. The origin of these SNIa is presumably in the stellar mass residing in the diffuse intracluster light (ICL). In a stacking analysis on several hundred clusters, \citet{Zibetti} constrained ICL within the 500 kpc extent from the cluster centre to be $10.9\pm 5.0$ per cent of the total cluster light. Other estimates are also provided in the literature, e.g., \citet{krick} found ICL between 6 - 22 per cent of the total cluster light within one quarter of the virial radius in optical r-band. Similarly, based on the SNIa rates, \citet{Dilday} estimated that about 1 - 3 per cent of the total mass may be residing in the diffuse intra-group light. Based on these observations, nearly 1 per cent of the total mass in the clusters is likely to be residing in the ICM as stars. Following the mass-normalized present-epoch SNIa rate provided in \citet{Dilday}, the IC-SNIa rate comes out to be $\sim$0.01 SN per year for diffuse IC stellar mass taken as 1 per cent of 10$^{14}$ M$_{\odot}$ of total cluster mass. We now discuss the possible roles that these SNIa can play in the ICM. 

\subsection{Filling of ICM with CR particles via IC-SNIa}

It can be examined now if the SNIa rates constrained by various observations are sufficient to fill the ICM with the synchrotron emitting CRe  within their typical lifetime of 100 Myr. We assume that the SNIa events take place more or less randomly in the ICM. The CRe are presumably accelerated in the SNe shock-waves for a short time ($\ll$1 Myr) and within that period, the SNe ejecta expands to less than 0.1 kpc radius in the ICM (see e.g., \citealt{Tang}). The CRe generated in this compact volume will diffuse into the ICM. The diffusion of CRe in the ICM is not well understood. \citet{Jaffe} suggested that the Alfv\'en velocity acts as an upper limit to the diffusion speed of CRe in the ICM. The CRe can therefore diffuse with a maximum rate of $\sim$0.1 kpc Myr$^{-1}$ at the Alfv\'en velocity of $\sim$100 km s$^{-1}$ in typical ICM environment. We consider the requirement of filling a total connected (in projection in the sky-plane) surface area ($\sim$10$^{5}$ kpc$^{2}$) as well as total volume ($\sim$10$^{8}$ kpc$^{3}$) corresponding to 500 kpc extent (diameter) through randomly occurring host-less SNIa events in the ICM within a 100 Myr period. The chosen extent of 500 kpc matches with the extents of the largest known RMH and also with typical extent often taken to estimate various observational properties of the massive clusters. With a rate of 0.01 SNIa per year, the cumulative surface area and volume filling through diffusion in 100 Myr will be $\sim$10$^{8}$ kpc$^{2}$ and $\sim$10$^{9}$ kpc$^{3}$ respectively. For mathematical convenience, we divided entire time period in bins of 100 yr and summed the filling contributions from the SNIa remnants through diffusion in each time-bin. 

The order of magnitude estimate made here indicates that a rate of 0.01 SNIa per year is sufficient with a good margin to collectively fill the ICM volume for 500 kpc extent with the CRe, generated in multiple (10$^{6}$) SNIa remnants over a 100 Myr period. These CRe are expected to emit diffuse synchrotron radio emission in the presence of magnetic field in the ICM. The excess Fe abundance in the CC clusters is seen up to nearly 100 kpc radius and the ICL is detected and constrained up to a few hundred kpc distance from the cluster center. Therefore, the filling of CRe via SNIa remnants cannot be arbitrarily extended to larger radii in the clusters. The CRe generated in the SNIa remnants can be expected to play a role in generating RMH, detected over a few hundred kpc scales in the central regions and may not be invoked to explain the diffuse radio emission in the giant radio halos seen over much larger scales. It is worth to mention here a possibility that if the particles accelerated in the SNIa remnants are primarily protons that can produce secondary electrons in the hadronic interactions with the thermal protons in the ICM \citep{Den}, the severity of this problem may get reduced to some extent due to the longer lifetime of the protons, which in turn enhances the possibility to fill a larger cluster volume.

\subsection{SNIa energy and radio powers of RMH}

Total energy deposited in the ICM due to the estimated number of $10^{6}$ IC-SNIa taking place in a 100 Myr period will be $10^{57}$ erg (revisable upwards, see below). This energy although much lower than the energy ($10^{60}$ erg; \citealt{Fujita}) input by the sloshing gas, is within the range of the energy in the X-ray cavities ($10^{55}-10^{61}$ erg; see \citealt{brunetti14}) created in AGN jets, which are also considered a potential source of energy in the ICM. Much of the total energy from the sources in the clusters is expected to go in to the heating of the ICM. In order to explain diffuse radio emission, energy rates in 100 Myr period is more relevant than the total energy deposition over cosmological time scales. The IC-SNIa with the rate of 0.01 per year will have average energy input rate of $3\times10^{41}$ erg s$^{-1}$. With the observed radio powers of the majority of RMH in the range $10^{39} - 10^{41}$ erg s$^{-1}$ \citep{Bravi}, total radio power emitted over 100 Myr period at their present rates will be $10^{54} - 10^{56}$ erg, which is lower than the total energy deposition by IC-SNIa in the same period. The maximum possible energy inputs by the cooling flows (mass deposition) are estimated in the range $10^{42} - 10^{45}$ erg s$^{-1}$ \citep{Bravi}. It is worth to point out the IC-SNIa rate will be a function of both cluster mass and a parameter called the delay time distribution (DTD) of SNIa which predicts the number of SNIa as a function of time ($t$) from the last star-burst. The DTD of the SNIa rate is predicted to be following a power-law of the form $\sim$1/$t$ between 0.1 and 10 Gyr (Maoz, Mannucci \& Brandt 2012).  In the sample of RMH of \citet{Gia14} the mass range of the clusters is $10^{14}-10^{15}$ M$_{\odot}$. Therefore, the assumed SNIa rate (i.e., 0.01 per year for $\sim$10$^{14}$ M$_{\odot}$) can vary up to 2 orders of magnitude depending upon the cluster mass and age (largely unknown) of the stellar population in the ICM. Consequently, the total energy deposited in the ICM due to IC-SNIa in a given 100 Myr period can also vary by the same factors and can reach up to $10^{59}$ erg. Therefore, the required energy conversion efficiency ($E_{\mathrm{CR}}/E_{\mathrm{SNe}}$) to match the observed radio powers of the detected RMH is expected to be $\le1$ per cent in the most favourable scenario. This efficiency is larger than that required to generate CRe in turbulent re-acceleration models extracting energies from gas-sloshing and cooling flows. The energy conversion efficiency for CRp in SNe explosions is estimated to be 2.8 per cent \citep{Berez}. Therefore, in the scenario involving proton acceleration in SNIa remnants and subsequent production of CRe in hadronic interactions \citep{Den}, the IC-SNIa can be a significant source of CRp throughout the cluster volume in the inner regions provided the CRp production in IC-SNIa is similar to that in the galactic environment. The impact of this additional spatially distributed source of CRp on diffuse radio emission from clusters needs to be assessed in future works. 

A rough estimate for the radio power emitted by CRe in shock accelerations (DSA process) using the formulation provided in \citet{Hoeft} can be made. We use physical parameters for a redshift $z\sim0.3$ cluster same as those used in the normalization (e.g., ICM magnetic field as 1$\mu$G, downstream plasma temperature as $7\times10^{7}$ K) of eq. 32 in \citet{Hoeft} except that the electron density is taken as 10$^{-3}$ cm$^{-3}$ relevant for cluster cores. The shock surface area per SN is estimated to be $\sim$0.02 kpc$^{2}$ for a spherical blast wave with radius 40 pc attained in the first 0.01 Myr time since the explosion in which the blast wave remains supersonic and follows the Sedov solutions in the ICM density 0.002 cm$^{-3}$ and temperature $5\times10^{7}$ K  (see \citealt{Tang}). For $10^{6}$ SNe, total shock surface area is estimated to be $2\times10^{4}$ kpc$^{2}$. The total radio power is predicted to be $\sim4\times10^{41}\Psi(\mathcal{M})$ erg s$^{-1}$. In order to match the observed radio powers of RMH, the shock strength parameter $\Psi(\mathcal{M})$ in \citet{Hoeft} needs to be between $10^{-2}$ and $1$. Again considering the two orders of magnitude uncertainties in the assumed SNIa rate, the predicted radio powers are revisable upwards depending upon the mass of the cluster and age of stellar population or $\Psi(\mathcal{M})$ can be revised downwards. The estimated radio power may also get scaled up in the cool-cores with higher electron density seen above 10$^{-2}$ cm$^{-3}$ such as in the Perseus cluster core (e.g., \citealt{Churazov}). Therefore, the required values for $\Psi(\mathcal{M})$ are expected to be small ($\ll1$), for which the corresponding $\mathcal{M}$ will be $3\pm1$ (see fig. 4 in \citealt{Hoeft}). The Mach numbers ($\mathcal{M}\approx(\upsilon/\mathrm{km~s^{-1}})~(T/100\mathrm{K})^{-0.5}$) are expected between 1 and 3 at the free expansion velocity ($\upsilon$) of the SNe ejecta in typical core ICM temperatures ($T=10^{7} - 10^{8}$~K). The $\mathcal{M}$ can be high (up to 10) for short periods ($<0.01$ Myr) in the early phases of the expansion of the SN remnant when the velocity of the ejecta is higher. The predicted radio powers can therefore be considered in close agreement with the observed radio powers of the RMH for the plausible values of $\mathcal{M}$ in the ICM and within the presently existing theoretical framework and various uncertainties.

\subsection{Observational consistencies}

There are a few interesting observational properties of the RMH in the clusters which are broadly consistent with the SNIa-scenario discussed in this paper. We briefly discuss those here.

\noindent(1) Two most powerful RMH are detected in the clusters RX J1347.5–1145 and MACS J1931.8-2634 \citep{Gia14}. The cluster RX J1347.5–1145 is among the most massive clusters known so far and shows bright diffuse stellar envelope around the BCG \citep{bradac}. The optical images of MACS J1931.8-2634 show intense diffuse ICL \citep{ehlert}. \citet{Gia17} found that the fraction of the clusters having RMH increases for the clusters above total mass of $M_{500\mathrm{kpc}} > 6 \times 10^{14}$ M$_{\odot}$. A direct implication of these observations could be that the IC stellar mass is sufficiently high in the high mass clusters thereby increasing the rate of IC-SNIa. 

\noindent(2)  Recently, \citet{Bravi} have argued in favour of a common origin for the gas heating and powering RMH via turbulence in the cool-cores. As SNIa are also viable sources of distributed heating and turbulence in the ICM, a role of SNIa in RMH can be indicated. 

\noindent(3) Based on a weak correlation between radio power of BCG and that of RMH, \citet{Gia14} suggested a possibility of a relation between nuclear activities in BCG and RMH. This relationship favours the SNIa scenario as both the nuclear activities in the BCG and IC-SNIa rate are possibly related to merger and star formation histories. It is assumed that nuclear activities and star formation are triggered by major merger events in the past. As the SNIa rate drops with time since the last major starburst activities with a power law (\citealt{Maoz}), it is essential that the last epoch of major star formation is not too old (a few Gyr) in order to sustain an adequate SNIa rate in the ICM. In that case, the radio powers of the RMH and possibly also the detections are very likely linked to the factors affecting global star formation in clusters such as the merger history. 

\noindent (4) A positive trend between radio power of RMH and X-ray luminosity has been reported \citep{bacchi, Kale15}, however, no clear trend between radio power of RMH and cluster temperature is found \citep{Gia14}. It implies that the radio powers of the RMH are closely related to the thermal electron density in the ICM. A particle acceleration mechanism such as that described in \citet{Hoeft}, which generates CRe from the available pool of seed thermal electrons, will fit into these observations. An analysis of radio emission in shock acceleration \citep{Hoeft} predicts a sharp decrease in the power of the radio emission with $\mathcal{M}$ below 5 with little possibility of radio emission below a critical $\mathcal{M}$ of $\sim$3. As $\mathcal{M}\propto T^{-0.5}$, it is very likely that the lower temperatures in the CC clusters assist to get the Mach numbers close to or above the critical value of $\sim$3 in the core and thereby making the particle acceleration efficient.

\section{Concluding remarks} 

A possibility of generating radio synchrotron emitting CRe in the shock fronts of the SNIa remnants in the ICM was discussed. The order of magnitude estimate made here indicates that a rate of 0.01 SNIa per year for a $10^{14}$ M$_{\odot}$ of the total cluster mass, is sufficient with a good margin to fill the ICM volume corresponding to 500 kpc extent with the CR particles, accelerated in randomly occurring multiple SNIa events over a 100 Myr period and diffusing with Alfv\'en velocity in the ICM. Total radio power emitted by RMH over a given 100 Myr period with a rate equal to the presently observed radio flux is lower than the total energy deposited by IC-SNIa in the ICM. Considering various uncertainties in the estimates, the required values for the energy conversion efficiency ($E_{\mathrm{CR}}/E_{\mathrm{SNe}}$) are moderate at a level of $\le1$ per cent in the most favourable scenario.  An estimate made for the predicted radio power in accelerations in $\mathcal{M}\sim$3 shocks in the IC-SNIa remnants is in good agreement with the observed radio powers of a good fraction of RMH. The relatively lower temperature in the CC clusters might be crucial in getting $\mathcal{M}$ near the critical number of 3 to get detectable radio emission from RMH. In addition, the high electron density in the cool-cores is also expected to increase the number density of the CR particles responsible for the diffuse radio emission. Several relations between RMH and CC clusters appear to be qualitatively consistent with the IC-SNIa scenario.  Although, the scenario discussed here is more suitable to explain radio halos in the CC clusters due to a possibility of having SNIa only in the inner a few hundred kpc extents as constrained from Fe abundance profiles in the CC clusters and detections of intra-cluster light in the ICM, it may have some applicability for explaining radio halos in general.

The most important aspect of the proposed scenario is that the IC-SNIa are distributed sources of turbulence and heating in the ICM and consequently also of the acceleration of the charged particles. The SNIa scenario requires only a moderate diffusion of CRe up to nearly 10 kpc in the lifetime of CRe in typical ICM magnetic field of 1 $\mu$G. The required energy conversion efficiencies for the SNIa scenario are somewhat larger compared to those in CRe re-acceleration in cooling-flow or gas-sloshing scenarios. A situation involving acceleration of protons in SNe remnants and subsequent production of secondary CRe in hadronic interactions is more attractive over solitary production of primary electrons in the SNe remnants. The impact of this possibility in producing radio emission via hadronic interaction path needs to be assessed in future works.

A presence or absence of RMH and also giant halos does not have a one to one relation with the merger stage inferred from cluster morphological state estimators \citep{Cassano, Kale15}. It then makes difficult to explain radio-ON and OFF stages of the clusters.  In high density ($\sim10^{-2}$ cm$^{-3}$) core such as in the Perseus cluster, the loss time-scales of the fossil electron population required in the re-acceleration models could be short ($\le1$ Gyr; \citealt{Sarazin,Pinzke}). As turbulent re-acceleration is sustainable on time-scales of $<1$ Gyr \citep{Hallman} from the time of injection of the turbulence, fresh injections of suprathermal/mildly-relativistic particles as well as turbulence are required to sustain the radio emission over periods $\ge1$ Gyr. The spatially and temporally distributed IC-SNIa remnants in the ICM can continuously inject a fresh population of electrons and turbulence required in the re-acceleration models.

The sample size of the known RMH is still statistically small and sensitivities of the radio images of RMH are not sufficient to look for some relationships between the radio power or morphology with the SNIa related properties of the clusters. Due to their dependence on the SNIa rate, the star formation and merger histories and metal abundance profiles of the CC clusters also need to be studied to find possible relationships between RMH power or morphology, SNIa rate and morphology of the ICL in the clusters. It is worth to mention that all the leptonic models presently discussed in the literature are qualitative at the moment. This paper is suggesting IC-SNIa as additional sources for CR particles in the ICM and generating diffuse radio emission in radio halos along with the other possibilities already discussed in the literature. At this stage, it does not attempt to rule out or favour one scenario over the other. In fact, multiple mechanisms may be at actions in the ICM for generating radio halos. The scenario presented in this paper needs to be validated, particularly in some numerical simulations.

\section*{Acknowledgements}

Author thanks the referee for prompt reviews of the manuscripts and giving insightful comments. The author acknowledges discussions with the students in the class-room teaching, which prompted to carry out the work presented in this paper.









\bsp	
\label{lastpage}
\end{document}